\documentclass[]{spie}  

\usepackage{amsmath,amsfonts,amssymb}
\usepackage{graphicx}
\usepackage[colorlinks=true, allcolors=blue]{hyperref}

\title{Exploring generative design AI tools for astronomical instrumentation: a CubeSat chassis case study}

\author[a]{Younes Chahid}
\author[a]{Tassos Aretos}
\author[a]{Will Cochrane}
\author[a]{Katherine Morris}
\author[a]{David Isherwood}
\author[a]{Zeshan Ali}
\author[a]{Graham Wilks}
\author[a]{Noah Schwartz}
\author[b]{Anmol Goyal}
\author[a]{Marcell Westsik}
\author[a]{Alastair Macleod}
\author[a]{Steve Watson}
\author[a]{Anjali Bhatt}
\author[a]{David Lunney}
\author[a]{Bradley Frank}

\affil[a]{STFC UK Astronomy Technology Centre, Royal Observatory Edinburgh, EH9 3HJ, United Kingdom}
\affil[b]{STFC Rutherford Appleton Laboratory, Harwell Campus, Didcot, OX11 0QX, United Kingdom}

\authorinfo{Send correspondence to younes.chahid@stfc.ac.uk}

\begin{document} 
\maketitle

\begin{abstract}
Generative design artificial intelligence (AI) tools are currently used in multiple scientific fields, yet their adoption in mechanical engineering computer-aided design (CAD) remains limited due to a lack of disseminated case studies, limited availability of accessible tools, insufficient training in CAD data, the absence of universal editable file formats and more. Mechanical design for astronomical instrumentation faces increasing complexity in thermal, vibrational, and mechanical requirements alongside tight project deadlines. This paper presents a practical evaluation of an AI and FEA based generative design tool applied to chassis design for the Active Deployable Optical Telescope (ADOT) CubeSat mission. Our analysis showcases the workflow steps including the setting of design, manufacturing and objective constraints. This study also shows the clear benefits of these types of tools, especially in the early brainstorming stages of multi-constrained mechanical structures, while also highlighting clear limitations like their black-box nature, the non-manufacturing-ready state of the results, and the time-consuming setup, limiting the tangible gain of these tools to high-value mechanical components.

\end{abstract}


\section{INTRODUCTION}
\label{sec:intro}  

CubeSat platforms have emerged as a cost-effective solution for space-based astronomical instrumentation, enabling increasingly sophisticated missions within highly constrained volumes and mass budgets. However, the integration of high-performance optical payloads within a CubeSat form factor introduces significant structural challenges. The supporting chassis must provide sufficient stiffness and stability to maintain optical alignment under launch loads and thermal variations, while remaining lightweight and compatible with manufacturing constraints.

Conventional mechanical design approaches rely on iterative computer-aided design CAD modelling and engineering intuition, which can limit exploration of the full design space. In recent years, advanced manufacturing methods\cite{chahid2024additive} and AI based generative design tools\cite{mcclelland2022generative,vlah2020evaluation,mcclelland2025evolved,buonamici2021generative} have been developed to automatically explore a wide range of geometries based on user-defined objectives, constraints, and load cases. These tools leverage optimisation algorithms to produce non-intuitive, high-performance structures, often achieving improved stiffness-to-weight ratios compared to conventional designs.

\begin{figure}
    \centering
    \includegraphics[width=0.7\linewidth]{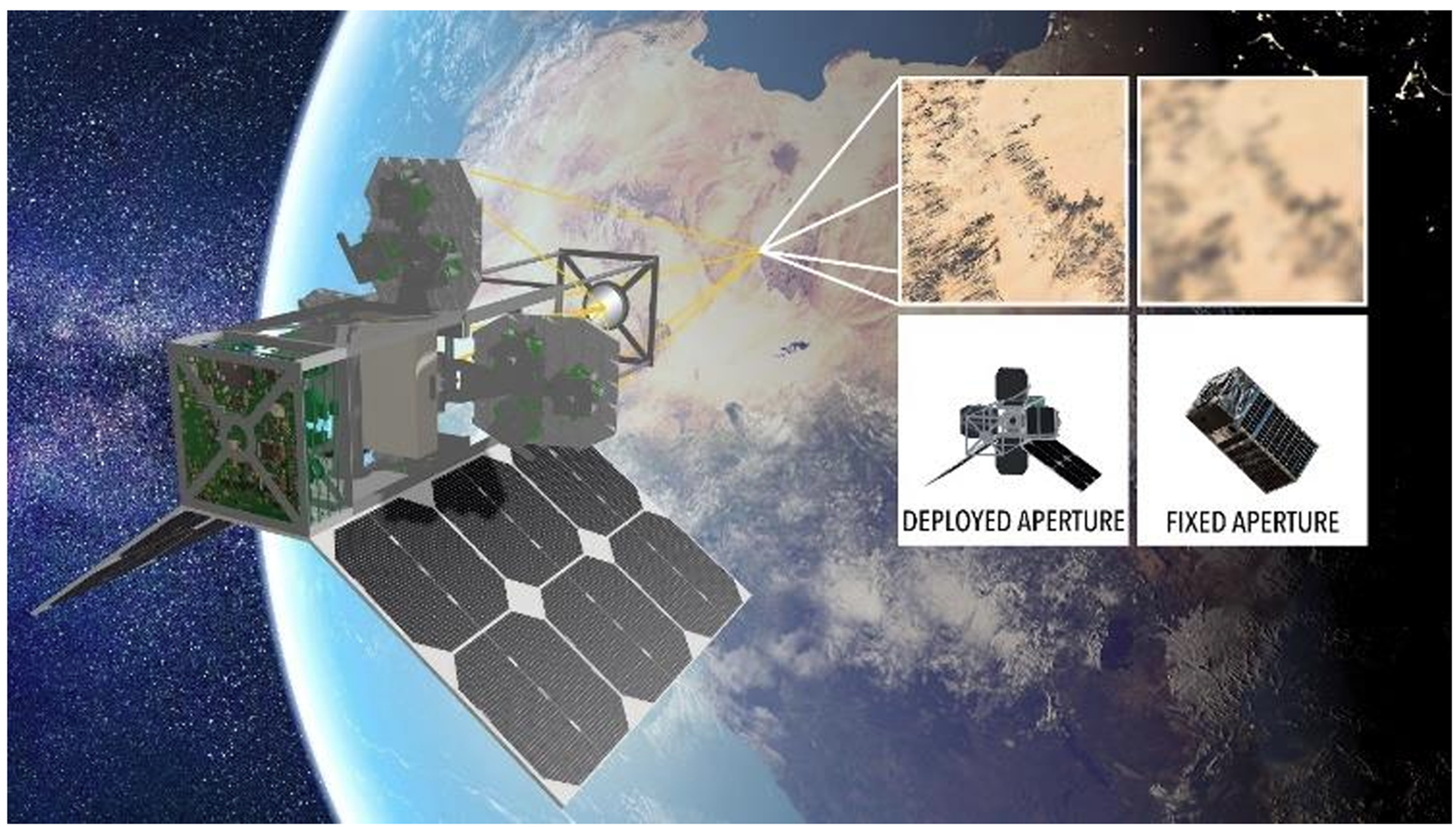}
    \caption{Illustration of the deployed  ADOT 6U CubeSat. Image credit: Schwartz, et al,. (2022).}
    \label{fig:space}
\end{figure}

This work investigates the application of generative design to the structural chassis of the Active Deployable Optical Telescope (ADOT), a CubeSat  mission concept featuring a deployable optical system \cite{schwartz20226u} seen in Fig. \ref{fig:space}. Using the generative design capabilities within Autodesk Fusion 360, the baseline chassis is re-designed with the objective of reducing mass while maximizing stiffness and and respecting multiple manufacturing constraints. The resulting designs are analyzed and the suitability and readiness of the workflow is qualitatively discussed alongside challenges and future work needed to increase its wider adoption.
The aim of this study is to evaluate the potential of generative design tools as part of the mechanical design workflow for astronomical instrumentation, and to highlight both the opportunities and limitations.

\section{Methodology}

\begin{figure}
    \centering
    \includegraphics[width=0.7\linewidth]{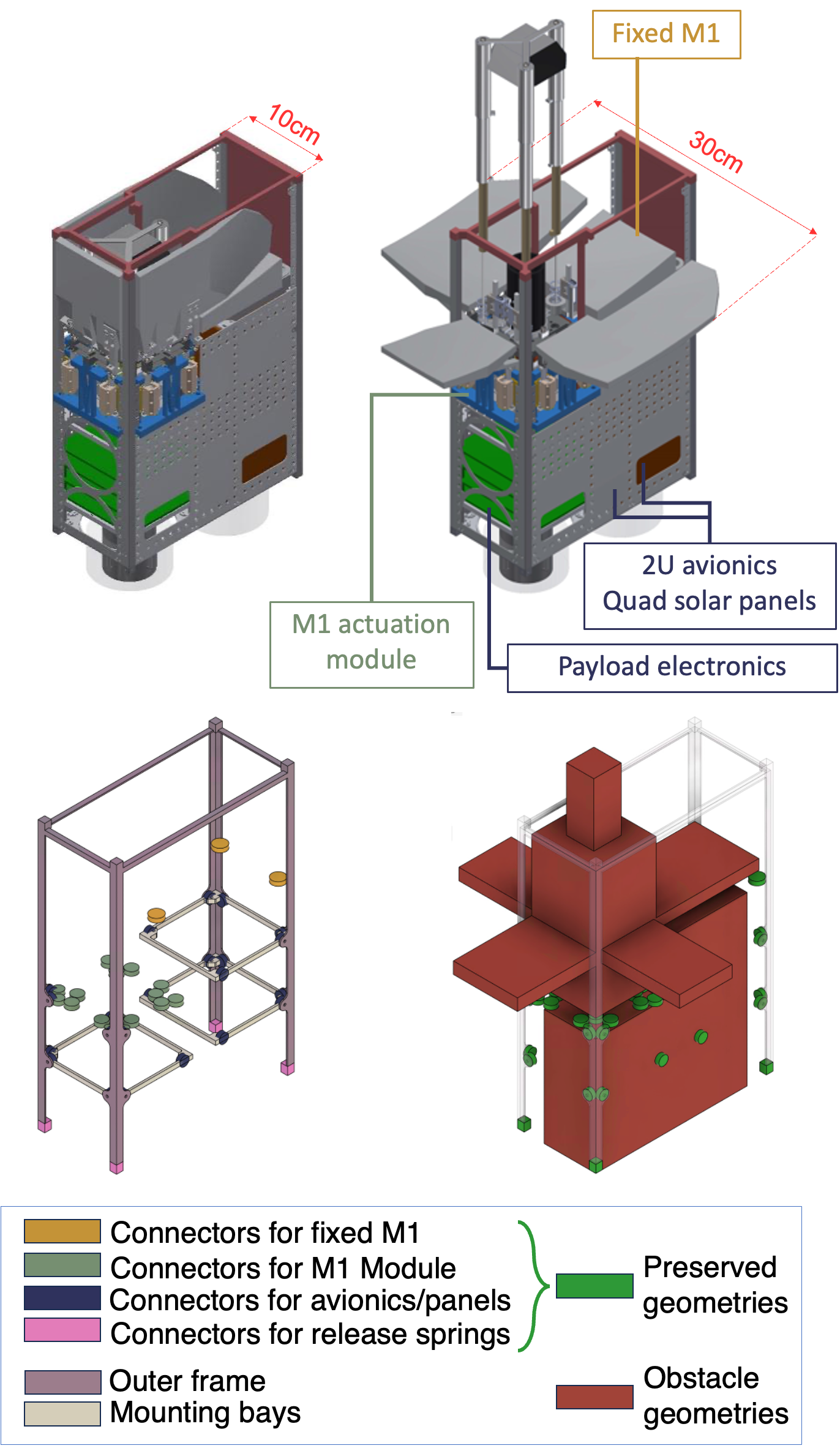}
    \caption{Stowed and deployed CubeSat\cite{morris2024accurate,morris2024field} configurations 
(Top image credit: Morris, et al,. (2024); Breakdown and simulation input geometry (down).}
    \label{fig:designspaces}
\end{figure}

ADOT 6U cubesat is composed of deployable components like the primary and secondary mirrors, allowing for a telescope aperture that is larger than the platform size. Detailed top level requirements can be found in previous publication \cite{schwartz20226u}. This study will use as a reference the conventional design\cite{morris2024accurate,morris2024field} shown in Fig. \ref{fig:designspaces}. 

\subsection{Design constraints}
\label{sec:title}

The AI based generative design process used in this study, seen in Fig.\ref{fig:workflow}, is an iterative FEA-driven process that can support a mechanical design engineer explore optimal conceptual geometries for a set of constraints defined as follow:

\begin{figure}
    \centering
    \includegraphics[width=0.7\linewidth]{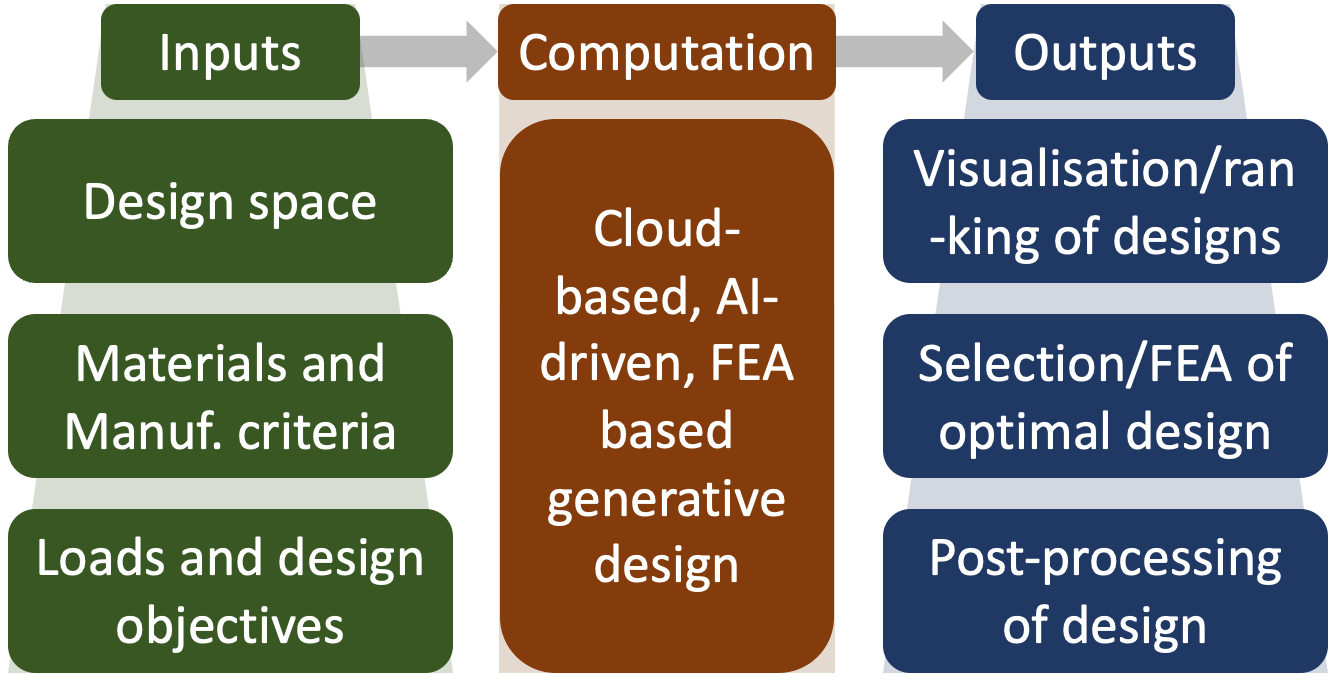}
    \caption{AI generative design process.}
    \label{fig:workflow}
\end{figure}

\begin{itemize}
    \item \textbf{Design Space:}
    \begin{itemize}
        \item \textit{Starting shape:} while not used in this study, this option allows the user to define a starting shape to iteratively lightweightof the FEA-based iterative lightweighting process. In our case, the starting shape was automatically generated by the software based on the preserved and obstacle geometries defined below.
        \item \textit{Preserved geometry:} highlighted in green in Fig. \ref{fig:designspaces}, these geometries represent what needs to be preserved as a solid geometry at the end of the lightweighting process. This list was defined as cylinders, representing the interfacing connectors needed for different payload components ranging from the M1 fixed mirror petal to the M1 module, solar panels and interface connectors for the release springs.
        \item \textit{Obstacle geometry:} highlighted in red in Fig. \ref{fig:designspaces}, this region will be avoided by the generative design algorithm. Obstacle regions in this case represent the payload in its deployed format. To avoid errors during the design generation step, it is important to have a direct line of sight between at least two or more preserved geometries. 
    \end{itemize}

    \item \textbf{Loads and Materials:}
    \begin{itemize}
        \item \textit{Structural Constraints:} fixed displacement constraints applied on `connectors for release springs' shown in Fig. \ref{fig:designspaces}.
        \item \textit{Structural Loads:} gravity and weight of payload parts applied on remaining connectors also shown in Fig. \ref{fig:designspaces}.
        \item \textit{Material:} three different aluminum alloys were chosen to adjust to the different explored manufacturing methods, in this case, aluminum alloy AlSi10Mg for additive manufacturing, Al6061 for machining and Al A536 T6 for casting.
    \end{itemize}

    \item \textbf{Manufacturing Criteria:}
    \begin{itemize}
        \item \textit{Additive manufacturing:} the use of print supports can be minimised by choosing an overhang angle of: 45$^\circ$ and a print axis orientation of: Y$+$. To ensure manufacturability, minimum material \ thickness was set to 5\,mm.
        \item \textit{Milling:} The chosen milling tool axis can be seen in \ref{fig:renders}. The milling tool parameters were also added as the diameter: 10\,mm, tool shoulder length: 40\,mm and head diameter: 60\,mm. These parameters would make sure that the tool can reach the generated geometry features, increasing the manufacturability of the generated design.
        \item \textit{Casting:} Ejection direction: $Z$, a minimum draft of angle: 3.0$^\circ$ and a minimum thickness: 5\,mm.
    \end{itemize}

    \item \textbf{Design Objectives:}
    \begin{itemize}
        \item Maximise stiffness.
        \item Mass target $<$ 1.5\,kg.
        \item Minimum firs modal frequency of 100\,Hz, based on NASA's Cubesat qualification recommendation \cite{GSFC-STD-7000B}.
        \item Symmetry in $XY$ plane (see Fig. \ref{fig:renders}).
    \end{itemize}
\end{itemize}

\section{Results}
\label{sec:sections}
\begin{figure}
    \centering
    \includegraphics[width=0.8\linewidth]{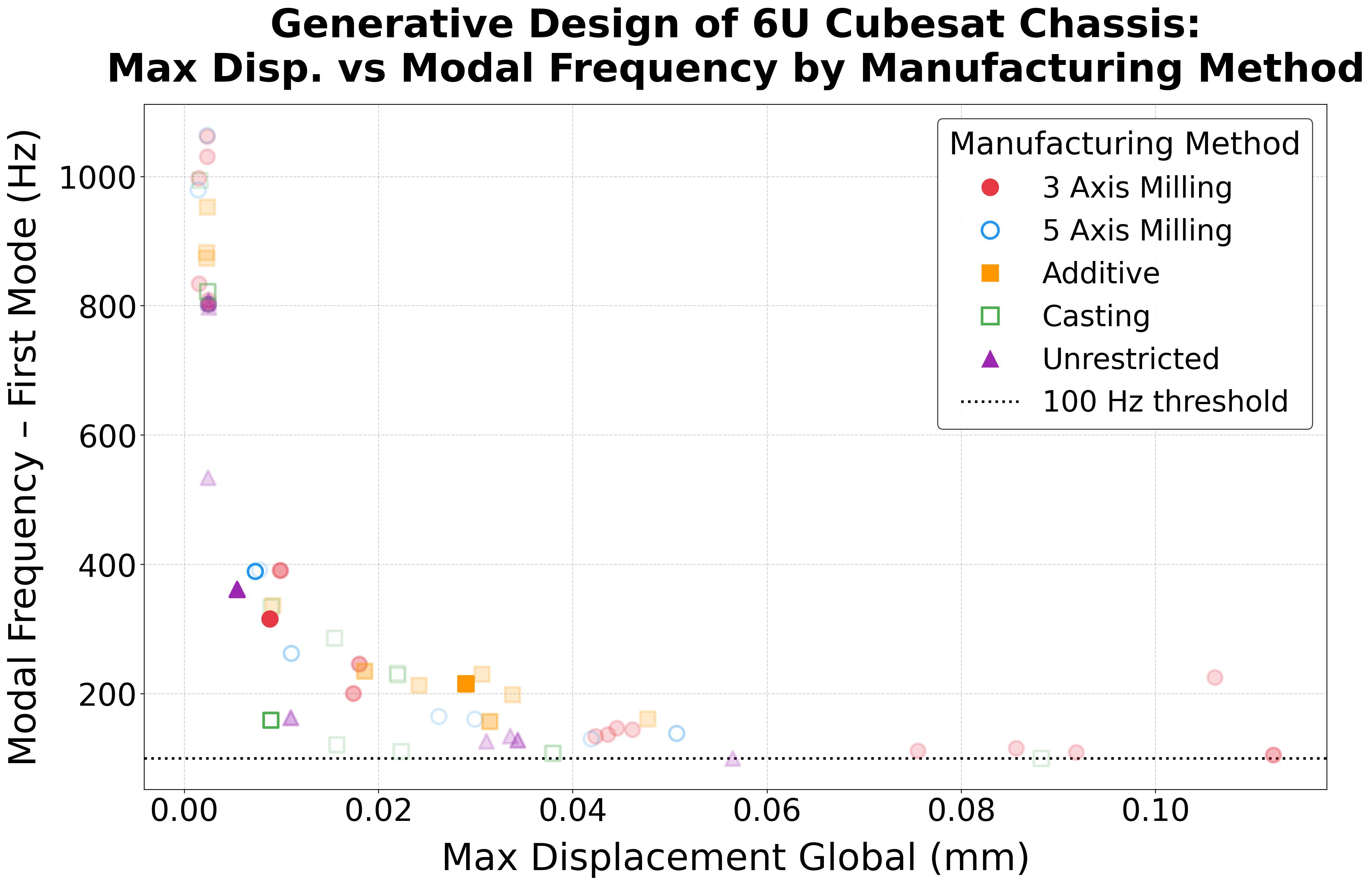}
    \caption{Maximum displacement versus first mode modal frequency for more than 80 generated designs. The opaque design values refer to the study seen in Fig. \ref{fig:renders}.}
    \label{fig:plot}
\end{figure}

\begin{figure}
    \centering
    \includegraphics[width=0.7\linewidth]{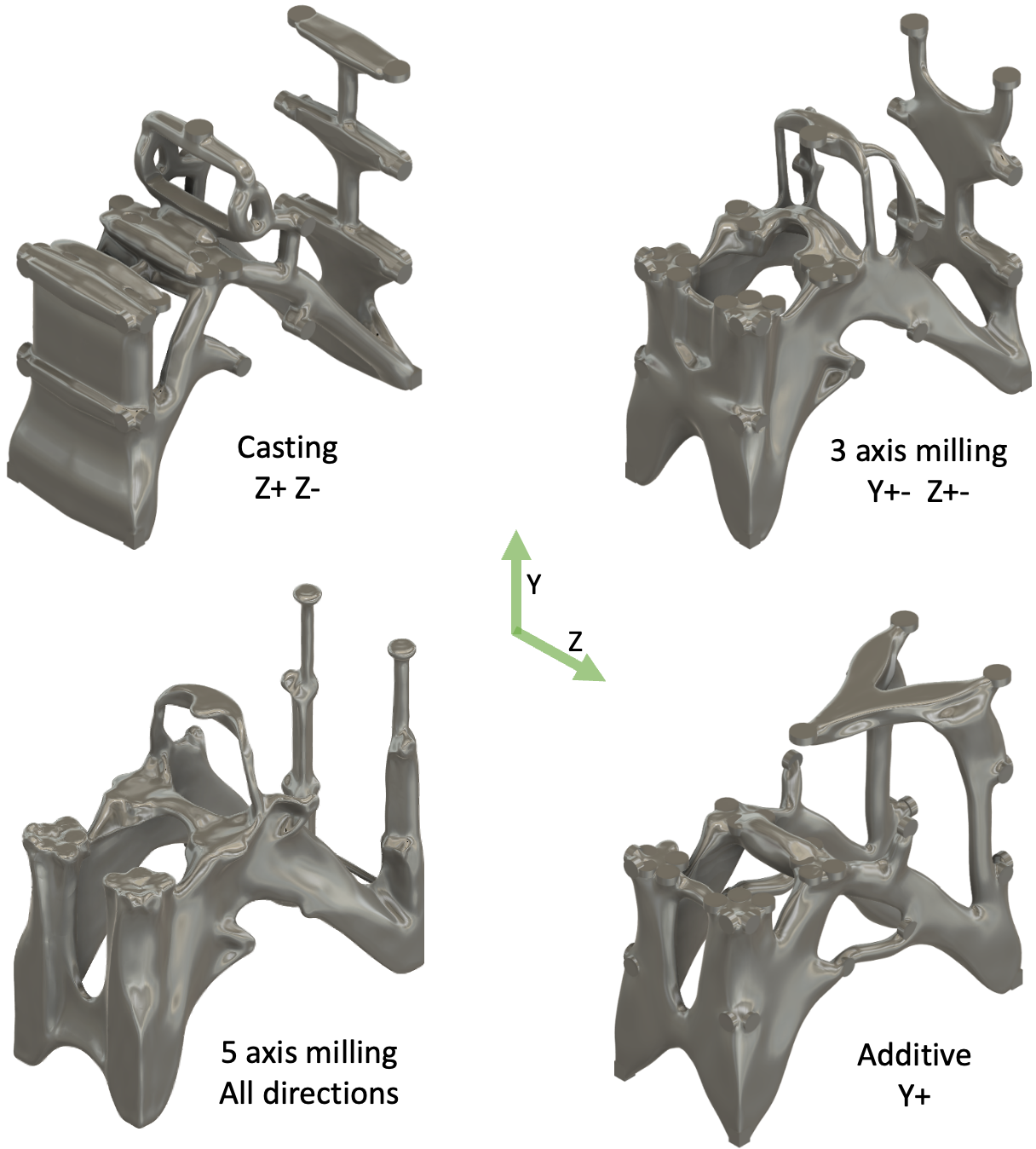}
    \caption{Example of constraints that maximise stiffness, mass target of 1.5\,kg and global displacement $< 0.1$\,mm.}
    \label{fig:renders}
\end{figure}

Because of the cloud-based solving, it was possible to obtain more than 80 generated designs in less than 2 hours. These results are plotted in Fig. \ref{fig:plot}. The high number of designs is due to the generative design algorithm generating a solution for each combination of material and corresponding manufacturing method. Additional design solutions were also generated by setting  different maximum global or local displacement values, further informing the generative design algorithm on the desired stiffness. For example, results shown in Fig.\ref{fig:renders} are specifically constrained to a global displacement $< 0.1$\,mm, on top of the above discussed design constraints. This generous global displacement value was deemed sufficient at this stage of the design process. Tighter displacement values will be evaluated in future studies in critical locations like the fixed M1 mirror and M1 module connectors.

Based on the plotted results seen in Fig. \ref{fig:plot}, all of the results respected the specific design of minimum first mode modal frequency of 100 Hz. However, from Fig. \ref{fig:renders}, other manufacturing constraints like overhang angle for the additive manufacturing design result was not respected, this can be attributed to the arrangement of the preserve and obstacle geometries, limiting the generative design algorithm in finding sufficient design space to meet this requirement.

In this workflow, the chosen geometry is decided by the mechanical design engineer, who will need to take in consideration the multiple design constraint parameters, the weight of each one, alongside other parameters like the availability of a manufacturing process, lead time and cost.

\begin{figure}
    \centering
    \includegraphics[width=0.7 \linewidth]{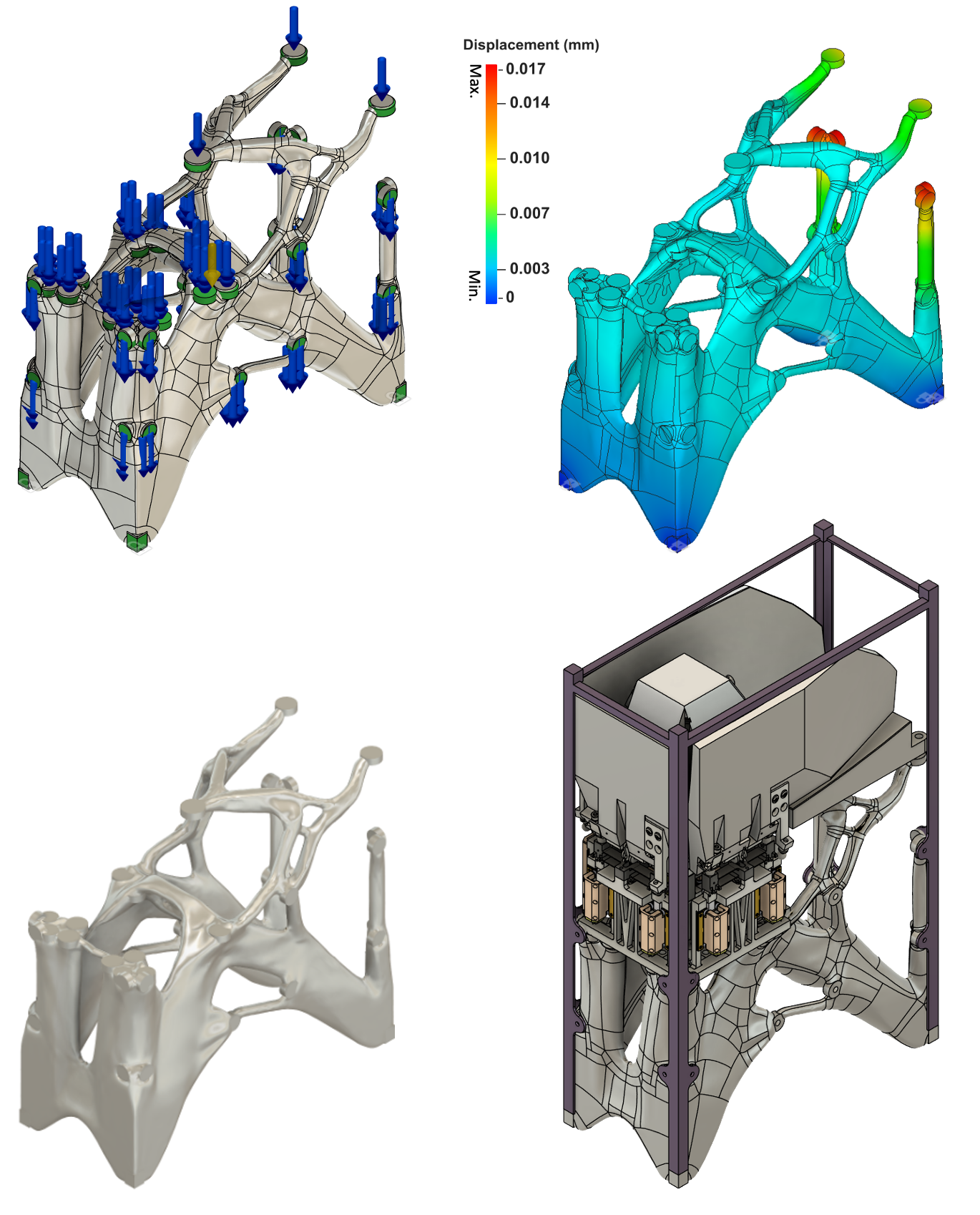}
    \caption{Structural loads and max displacement FEA simulation (up) applied on generated design (down).}
    \label{fig:FEA}
\end{figure}

In this study, and for demonstration purposes only, the additive manufacturing design option was chosen for the follow up post-processing steps, shown in Fig. \ref{fig:FEA}. The chosen geometry was exported to an editable CAD geometry. Follow up steps typically include a refinement of the design like adding holes or non-organic geometries that can simplify the mechanical drawings and the geometric dimensioning and tolerancing (GD\&T) of critical features. This step is followed by a re-application of similar FEA study and double checking of the obtained values, making sure the post-processed geometry results conform to the initial set of constraints. Lastly, the validated geometry was imported back to the CubeSat assembly for follow up mechanical design of interfaces Fig. \ref{fig:FEA}.

The following section will discuss the remaining challenges that needs to be tackled in order to allow a wider useful adoption of AI and FEA based generative design tools for mechanical design tasks.

\section{Discussion and conclusion}

The fast generation of more than 80 designs in multiple manufacturing constraints and materials can be very useful especially at the early design brainstorming phase of a project, initial selection of manufacturing process and rough costing.

However, despite these benefits, the obtained geometries 
are far from being ready for manufacturing and would still require design iterations in accordance to the chosen manufacturing method. For example, the addition of work-holding or sacrificial features used in the manufacturing or measurement phase for machining and casting or further analysis of print orientation and supports generation in the case of metal additive manufacturing. To tackle this challenge, future study will investigate the possibility of adding design for manufacturing preserved geometries that could improve the manufacturability and metrology process of the obtained geometry and accelerate its manufacturing process.

Another limitation of this process is its black-box nature, not only due to its implementation of AI algorithms but primarly due to the protective commercial nature of this study's chosen chosen algorithm. This limitation can restrict the user's understanding of the generative design process, therefore limit ways to quality control the obtained results or improve the process efficiency. To tackle this challenge, future work will look at investigating open-source AI and FEA-based mechanical generative design algorithms.

Lastly, the setup time of these types of studies can be, at least for now, time-consuming, which means tangible gains are mostly visible in mechanical parts that are critical and have a higher value. This means the estimated value gained by lightweighting or consolidating mechanical parts will have to be weighed against the time needed to set up and complete the generative design study, to justify the worth of undertaking it.

In conclusion, this study shows the clear benefit of generative design tools during the brainstorming phase. It also highlights their limitations in wider adoption within the mechanical design workflow. Further studies are needed to simplify the adoption process, increase transparency, and support the mechanical design engineer beyond the brainstorming phase. This includes helping decide whether the tool is worth using for a given part, avoiding black-box algorithms when possible, and assisting with further steps such as choosing the ideal design based on parameters beyond the design itself, like process specific hidden cost. Further tangible benefits in this field would also come from supporting post mechanical design steps such as GD\&T, mechanical drawings, and metrology measurement strategies.

\acknowledgments 
The authors gratefully acknowledge STFC’s Centre for Innovation (CfI) funding.

\bibliography{report} 
\bibliographystyle{spiebib} 

\end{document}